\documentclass[smallextended]{svjour3}       % onecolumn (second format)
\smartqed  % flush right qed marks, e.g. at end of proof
\usepackage{graphicx,epsfig}
\journalname{Quantum Inf. Process.}

\begin{document}

\title{
On the quantum correlations in two-qubit XYZ spin chains with Dzyaloshinsky-Moriya and
Kaplan-Shekhtman-Entin-Wohlman-Aharony interactions
}
\author{
M.A.Yurischev
}
\institute{
M.~A.~Yurischev
\at
Institute of Problems of Chemical Physics, Russian Academy of Sciences,
Chernogolovka 142432, Moscow Region, Russia\\     
\email{yur@itp.ac.ru} }

\date{Received:}

\titlerunning{
On the quantum correlations in two-qubit XYZ spin chains
}
\maketitle

\begin{abstract}
The anisotropic Heisenberg two-spin-1/2 model in an inhomogeneous magnetic field with
both antisymmetric Dzyaloshinsky-Moriya and symmetric
Kaplan-Shekhtman-Entin-Wohlman-Aharony cross interactions is considered at thermal
equilibrium.
Using a group-theoretical approach, we find fifteen spin Hamiltonians and as many
corresponding Gibbs density matrices (quantum states) whose eigenvalues are expressed
only through square radicals.
We also found local unitary transformations that connect nine of this fifteen state
collection, and one of them is the X quantum state.
Since such quantum correlations as quantum entanglement, quantum discord, one-way
quantum work deficit, and others are known for the X state, this allows to get the
quantum correlations for any member from the nine state family.
Further, we show that the remaining six quantum states are separable, that they are
also connected by local unitary transformations, but, however, now the case with known
correlations beyond entanglement
is generally not available.
\end{abstract}

%\PACS{03.65.Aa \and 03.65.Ta \and 03.67. -a \and 89.70.Cf}
% \and 42.50. -p}
%
%\subclass{Primary 81P40 \and Secondary 81Qxx}

%======================================================================
\section{Introduction}
\label{sect:Intro}
To explain the phenomenon of weak ferromagnetism observed in some rhombohedral
antiferromagnets, Dzyaloshinsky \cite{D57,D58} developed a phenomenological approach
based on the Landau theory of second-order phase transitions and showed that the
{\em antisymmetric} mixed (in magnetization components) term in the expansion of the
thermodynamic potential is responsible for the appearance of nonzero net magnetization
of the system.
Shortly after \cite{D57a}, he also noticed that in antiferromagnetic crystals with the
tetragonal lattices, weak ferromagnetism can be caused by the {\em symmetric} mixed
term in the expansion of the corresponding thermodynamic potential.

Later on, in 1960, Moriya \cite{M60,M60a} developed a microscopic theory of
anisotropic superexchange interaction by extending the Anderson
theory of superexchange to include spin-orbit coupling.
Using perturbation theory he found that the leading anisotropy contribution to the
interaction between two neighboring spins ${\vec\sigma}_1$ and ${\vec\sigma}_2$ is
given by
\begin{equation}
   \label{eq:H_DM}
   {\cal H}_{\rm DM}={\bf D}\!\cdot\!({\vec\sigma}_1\times{\vec\sigma}_2),
\end{equation}
where ${\bf D}=(D_x,D_y,D_z)$ is a constant vector that characterizes a substance.
This interaction reproduces Dzyaloshinsky's antisymmetric term and is now
referred to as the Dzyaloshinsky-Moriya (DM) interaction.
In addition, Moriya found the second-order correction term
\cite{M60,M60a}
\begin{equation}
   \label{eq:H_KSEA}
   {\cal H}_{\rm KSEA}={\vec\sigma}_1\!\cdot\!{\tilde{\rm\Gamma}}\!\cdot\!{\vec\sigma}_2,
\end{equation}
where ${\tilde{\rm\Gamma}}$ is a symmetric traceless tensor.
For a long time this interaction was assumed to be negligible compared with the
antisymmetric contribution (\ref{eq:H_DM}).
However, more later Kaplan \cite{K83} and then Shekhtman, Entin-Wohlman, and Aharony
\cite{SEA92,SEA93} argued the importance of the symmetric term because it can restore
the $O(3)$ invariance of the isotropic Heisenberg system which is broken by the DM
term.
For this reason, the interaction (\ref{eq:H_KSEA}) began to be called the
Kaplan-Shekhtman-Entin-Wohlman-Aharony (KSEA) interaction \cite{Zh98,Zh98a} (see also
reference 14 in \cite{YHAE95}).

We will discuss two-site systems with the Hamiltonian
\begin{equation}
   \label{eq:H}
   {\cal H}={\cal H}_{\rm Z} + {\cal H}_{\rm H} + {\cal H}_{\rm DM} + {\cal H}_{\rm KSEA},
\end{equation}
where ${\cal H}_{\rm Z}$ is the Zeeman energy and ${\cal H}_{\rm H}$ the
anisotropic exchange Heisenberg interactions.
Behavior of quantum correlations in different particular cases of the model
(\ref{eq:H}) was considered in numerous papers.
The behavior of thermal entanglement in two-qubit completely isotropic (XXX)
Heisenberg chain
in the absence of an external field but in the
presence of DM interaction with a nonzero of only one, $D_z$, component of the
Dzyaloshinsky vector ${\bf D}$ was considered in \cite{Z07}.
The author of this paper found that the DM interaction can excite entanglement.
Thermal entanglement in the partially anisotropic (XXZ) Heisenberg model with $D_z$ or
$D_x$ component of Dzyaloshinsky vector was studied in \cite{LWC08}.
Quantum entanglement in anisotropic Heisenberg XXZ chain with only
$D_z$ component was also discussed in \cite{KJL09}.
The authors established that while the anisotropy suppresses the entanglement the
DM interaction can restore it.
The effect of DM interaction on the quantum entanglement in the Heisenberg XYZ chain
was observed in \cite{LC09} in the absence of magnetic field.
Thermal quantum discord in the anisotropic Heisenberg XXZ model with DM interaction
and without any external field was investigated in \cite{CY10}.
Concurrence and quantum discord in two-qubit anisotropic Heisenberg XXZ model with DM
interaction along the $z$-direction was considered in \cite{TAMAP13} where it was
found that the tunable parameter $D_z$ may play a constructive role to the quantum
correlations in thermal equilibrium.
Quantum discord of two-qubit anisotropy XXZ Heisenberg chain with DM interaction under
uniform magnetic field was investigated in \cite{Z14}.
In the recent papers \cite{P19,SMG20}, the thermal quantum entanglement and
discord in two-qubit XYZ chain with DM interaction were discussed.
As a whole, one can conclude that the most results have been obtained for the spin
pairs with DM interactions when the exchange Heisenberg couplings are isotropic or,
rarer, anisotropic.
The external magnetic field was taken into account much less frequently.
Finally, there are no publications where the quantum correlations (entanglement,
discord, etc.) in Heisenberg dimers were discussed in the presence of KSEA
interactions.
Our research fills these gaps to a certain extent.

The structure of this paper is as follows.
In the next section, we write down the Hamiltonian in an expanded form and establish
its relationship with the density matrix.
Sects.~3 and 4 deal with the X and CS quantum states what gives the key that opens the
way, first, to the group-theoretical analysis in Sect.~5 and then, in Sect.~6, to the
results for the $D_y$ and ${\rm\Gamma}_y$ pair components of Dzyaloshinsky vector and
$\hat{\rm\Gamma}$ tensor.
Sect.~7 is devoted to the classification of fifteen Hamiltonians and density matrices.
Finally, in the last section~8, we briefly summarize the results obtained and note the
remaining unsolved problems.

%======================================================================
%\clearpage
%\newpage
\section{
Hamiltonian and density matrix
}
\label{sect:H-rho}
The DM interaction (\ref{eq:H_DM}) can be written in an expanded form as
\begin{equation}
   \label{eq:H_DM_Dxyz}
   {\bf D}\!\cdot\!({\vec\sigma}_1\times{\vec\sigma}_2)
	 =D_x(\sigma_1^y\sigma_2^z-\sigma_1^z\sigma_2^y)
	 +D_y(\sigma_1^z\sigma_2^x-\sigma_1^x\sigma_2^z)
	 +D_z(\sigma_1^x\sigma_2^y-\sigma_1^y\sigma_2^x),
\end{equation}
where ${\vec\sigma}_i$ denotes the vector of Pauli matrices at site $i=1,2$;
${\vec\sigma}_i=(\sigma_i^x,\sigma_i^y,\sigma_i^z)$.
Similarly for the KSEA term (\ref{eq:H_KSEA}):
\begin{eqnarray}
   \label{eq:H_KSEA_Gxyz}
	 &&{\vec\sigma}_1\!\cdot\!{\tilde{\rm\Gamma}}\!\cdot\!{\vec\sigma}_2
	=(\sigma_1^x,\sigma_1^y,\sigma_1^z)
	 \left(
      \begin{array}{lll}
      0&{\rm\Gamma}_z&{\rm\Gamma}_y\\
      {\rm\Gamma}_z&0&{\rm\Gamma}_x\\
      {\rm\Gamma}_y&{\rm\Gamma}_x&0
      \end{array}
   \right)\!
	 \left(
      \begin{array}{c}
      \sigma_2^x\\
      \sigma_2^y\\
      \sigma_2^z
      \end{array}
   \right)
	 \nonumber\\
	 &&={\rm \Gamma}_x(\sigma_1^y\sigma_2^z+\sigma_1^z\sigma_2^y)
	 +{\rm \Gamma}_y(\sigma_1^z\sigma_2^x+\sigma_1^x\sigma_2^z)
	 +{\rm \Gamma}_z(\sigma_1^x\sigma_2^y+\sigma_1^y\sigma_2^x),
\end{eqnarray}
where ${\rm \Gamma}_x$, ${\rm \Gamma}_y$, and ${\rm \Gamma}_z$ are the elements of the
tensor ${\tilde{\rm\Gamma}}$.
As a result, the Hamiltonian (\ref{eq:H}) is rewritten as
\begin{eqnarray}
   \label{eq:HHH}
   {\cal H}&=&B_1^x\sigma_1^x + B_1^y\sigma_1^y + B_1^z\sigma_1^z + B_2^x\sigma_2^x + B_2^y\sigma_2^y + B_2^z\sigma_2^z
	 \nonumber\\
	 &+& J_x\sigma_1^x\sigma_2^x + J_y\sigma_1^y\sigma_2^y + J_z\sigma_1^z\sigma_2^z
	 \nonumber\\
	 &+& D_x(\sigma_1^y\sigma_2^z-\sigma_1^z\sigma_2^y)
	 +D_y(\sigma_1^z\sigma_2^x-\sigma_1^x\sigma_2^z)
	 +D_z(\sigma_1^x\sigma_2^y-\sigma_1^y\sigma_2^x)
	 \nonumber\\
	 &+& {\rm \Gamma}_x(\sigma_1^y\sigma_2^z+\sigma_1^z\sigma_2^y)
	 +{\rm \Gamma}_y(\sigma_1^z\sigma_2^x+\sigma_1^x\sigma_2^z)
	 +{\rm \Gamma}_z(\sigma_1^x\sigma_2^y+\sigma_1^y\sigma_2^x),
\end{eqnarray}
where $B_i^\alpha$ ($i=1,2$; $\alpha=x,y,z$) are the components of the external
magnetic fields ${\bf B}_1$ and ${\bf B}_2$ (with the incorporated gyromagnetic ratios
or $g$-factors) and $J_\alpha$ ($\alpha=x,y,z$) are the Heisenberg exchange couplings.

We will be able to study the systems only in some special cases of the Hamiltonian
(\ref{eq:HHH}).
For them, we will be interested in systems in a state of thermal equilibrium.
The corresponding Gibbs density matrix is given as
\begin{equation}
   \label{eq:rhoG}
   \rho=\frac{1}{Z}\exp(-{\cal H}/T),
\end{equation}
where $T$ is the temperature in energy units and $Z$ the partition function.
Thus, the Hamiltonian and density matrix of any system are connected via the
functional relation.

Quantum correlations contained in composite quantum states are the focus of quantum
information science.
Many measures have been proposed to quantify these correlations,
such as quantum entanglement, quantum discord,
one-way quantum work deficit and so on
\cite{AFOV08,HHHH09,CMS11,MBCPV12,AFY14,ABC16,FPA17,BDSRSS18}.
It should be emphasized that the quantum correlation measures must satisfy a number of
criteria \cite{BM12} (see also the review \cite{MBCPV12}).
In particular, as a necessary condition, the measures must be invariant under any
{\em local\/} unitary transformations.

%======================================================================
%\clearpage
%\newpage
\section{
XYZ chain with $D_z$ and ${\rm\Gamma}_z$ couplings
}
\label{sect:X_Dz}
We begin the analysis with quantum states having the X form.
In accord with definition, X matrix can have nonzero entries only on the main diagonal
and anti-diagonal.
The portrait of such a sparse matrix resembles the letter ``X'', which allowed to give
it such a name \cite{YE07}.
Algebraic characterization of X states in quantum information has been done by Rau
\cite{R09}.
It is important to note that both the sums and the products of X matrices are again
the X matrices, that is, the set of X matrices is algebraically closed.
In particular, a function (decomposable in a Taylor series) of X matrix is the X
matrix.

In the most general form, the Hermitian X matrix corresponding to the Hamiltonian
(\ref{eq:HHH}) can be written as
\begin{equation}
   \label{eq:HXzz}
   {\cal H}_{zz}=
	 \left(
      \begin{array}{cccc}
      J_z+B_1^z+B_2^z&.&.&J_x-J_y-2i{\rm \Gamma}_z\\
      .&-J_z+B_1^z-B_2^z\ &J_x+J_y+2iD_z&.\\
      .&J_x+J_y-2iD_z\ &-J_z-B_1^z+B_2^z&.\\
      J_x-J_y+2i{\rm \Gamma}_z&.&.&J_z-B_1^z-B_2^z
      \end{array}
   \right),
\end{equation}
where the points are put instead of zero entries.
In Eq.~(\ref{eq:HXzz}), $B_1^z$ and $B_2^z$ are the $z$-components of external fields
applied at the 1-st and 2-nd qubits respectively, ($J_x$,$J_y$,$J_z$) the vector of
interaction constants of the Heisenberg part of interaction, $D_z$ the $z$-component
of Dzyaloshinsky vector, and ${\rm\Gamma}_z$ the $z$-component in the KSEA
interaction.
Thus, this model contains seven real independent parameters: $B_1^z$, $B_2^z$, $J_x$,
$J_y$, $J_z$, $D_z$, and ${\rm\Gamma}_z$.

On the other hand, ``any four-by-four matrix--and, therefore, the Hamiltonian matrix
in particular--can be written as a linear combination of the sixteen double-spin
matrices'' \cite{FLS64}, Sect.~12-2.
For the traceless X matrix (\ref{eq:HXzz}), the linear combination of
``double-spin matrices'' is given as
\begin{equation}
   \label{eq:Hzz}
   {\cal H}_{zz}=B_1^z\sigma_1^z + B_2^z\sigma_2^z
	 + J_x\sigma_1^x\sigma_2^x + J_y\sigma_1^y\sigma_2^y + J_z\sigma_1^z\sigma_2^z
	 + J_{xy}\sigma_1^x\sigma_2^y + J_{yx}\sigma_1^y\sigma_2^x),
\end{equation}
where 
\begin{equation}
   \label{eq:JxyJyx}
   J_{xy}=D_z + {\rm\Gamma}_z,\qquad J_{yx}=-D_z + {\rm\Gamma}_z.
\end{equation}

Due to the functional relation (\ref{eq:rhoG}), the Gibbs density matrix also has the
X form with seven real parameters:
\begin{eqnarray}
   \label{eq:rho-zz}
   \rho_{zz}&=&
	 \left(
      \begin{array}{cccc}
      a&.&.&u\\
      .&b\ &v&.\\
      .&v^*\ &c&.\\
      u^*&.&.&d
      \end{array}
   \right)
	 =\frac{1}{4}(\sigma_0\otimes\sigma_0 + s_1^z\sigma_z\otimes\sigma_0 + s_2^z\sigma_0\otimes\sigma_z
	 \nonumber\\
	 &+& c_1\sigma_x\otimes\sigma_x + c_2\sigma_y\otimes\sigma_y + c_3\sigma_z\otimes\sigma_z
	 + c_{12}\sigma_x\otimes\sigma_y + c_{21}\sigma_y\otimes\sigma_x),\
\end{eqnarray}
where the asterisk denotes complex conjugation,
$s_1^z$, $s_2^z$, $c_1$, $c_2$, $c_3$, $c_{12}$, and $c_{21}$ are the
unary and binary correlation functions, and
\begin{equation}
   \label{eq:s_0xyz}
   \sigma_0=
	 \left(
      \begin{array}{rr}
      1&0\\
      0&1
      \end{array}
   \right),\quad
   \sigma_x=
	 \left(
      \begin{array}{rr}
      0&1\\
      1&0
      \end{array}
   \right),\quad
   \sigma_y=
	 \left(
      \begin{array}{rr}
      0&-i\\
      i&0
      \end{array}
   \right),\quad
   \sigma_z=
	 \left(
      \begin{array}{rr}
      1&0\\
      0&-1
      \end{array}
   \right)
\end{equation}
are the unit and Pauli spin operators in the standard representation.
Due to the nonnegativity definition and normalization condition of any density
operator, $a,b,c,d\ge0$, $a+b+c+d=1$, $ad\ge|u|^2$, and $bc\ge|v|^2$.

One can now calculate different quantum correlations in the X quantum states.
The methods of calculating quantum correlations for the two-qubit X quantum states
has been developed in a number works.
The concurrence, a measure of quantum entanglement, is given by \cite{YE07}
\begin{equation}
   \label{eq:Conc}
   C=2\max\{0,|u|-\sqrt{bc},|v|-\sqrt{ad}\}.
\end{equation}
There are considerable studies on the quantum discord and one-way quantum work deficit.
For instance, the quantum discord of two-qubit X quantum states was considered in
Refs.~\cite{CZYYO11,H13,JY16} (and references therein).
One may present the quantum discord as a formula
\begin{equation}
   \label{eq:Q3}
   Q=\min\{Q_0, Q_{\tilde\theta}, Q_{\pi/2}\},
\end{equation}
where the subfunctions (branches) $Q_0$ and $Q_{\pi/2}$ are the analytical expressions
(corresponding to the discord with optimal measurement angles 0 and $\pi/2$,
respectively) and only the third branch $Q_{\tilde\theta}$ requires one-dimensional
searching of the optimal state-dependent measurement angle $\tilde\theta\in(0,\pi/2)$
(details see in Refs.~\cite{Y14,Y14a,Y15,Y17}).

Very similar situation takes place for the one-way quantum work deficit
\cite{WJFWCF15,YF16,YWF16}.
We may again write the one-way quantum work deficit of two-qubit X state in a
semi-analytical form:
\begin{equation}
   \label{eq:D3}
   {\rm\Delta}=\min\{\Delta_0, \Delta_{\vartheta}, \Delta_{\pi/2}\},
\end{equation}
where the branches $\Delta_0$ and $\Delta_{\pi/2}$ are known in the analytical form
while the third branch $\Delta_{\vartheta}$ also requires to perform numerical
minimization to obtain state-dependent minimizing polar angle $\vartheta\in(0,\pi/2)$
(see \cite{YWF16,Y18,Y19,Y19a}). 

So, the theory to calculate quantum correlations of X quantum states is well
developed.
This gives a possibility to calculate and investigate different quantum correlations
for the two-qubit systems in a nonuniform field in $z$-direction, with completely
anisotropic Heisenberg interactions, and with arbitrary $z$-components of DM and KSEA
interactions.

%======================================================================
%\clearpage
%\newpage
\section{
XYZ chain with $D_x$ and ${\rm\Gamma}_x$ couplings
}
\label{sect:CS_Dx}
Let us take the centrosymmetric (CS) quantum state now.
The CS matrix $n\times n$ is defined by the relations for its matrix elements as
follows: $a_{ij}=a_{n+1-i,n+1-j}$ \cite{W85}.
It is easy to check, that the sum and product of CS matrices are the CS matrix, i.e.,
this family of matrices as well as X matrices is algebraically closed.

Most general Hermitian CS matrix of fourth order looks as
\begin{equation}
   \label{eq:rho-xx}
   \rho_{xx}=
	 \left(
      \begin{array}{lccl}
      a&\mu&\nu&c\\
      \mu^*&b&d&\nu^*\\
      \nu^*&d&b&\mu^*\\
      c&\nu&\mu&a
      \end{array}
   \right),
\end{equation}
where $a$, $b$, $c$, and $d$ are real quantities while  $\mu$ and $\nu$ are complex.

The Hamiltonian with CS symmetry reads
\begin{equation}
   \label{eq:HCSxx}
   {\cal H}_{xx}=
	 \left(
      \begin{array}{cccc}
      J_z&B_2^x-iJ_{zy}\ &B_1^x-iJ{yz}&J_x-J_y\\
      B_2^x+iJ_{zy}&-J_z&J_x+J_y&B_1^x+iJ_{yz}\\
      B_1^x+iJ_{yz}&J_x+J_y&-J_z&B_2^x+iJ_{zy}\\
      J_x-J_y&B_1^x-iJ_{yz}\ &B_2^x-iJ_{zy}&J_z
      \end{array}
   \right).
\end{equation}
This Hamiltonian in the Bloch form is written as
\begin{equation}
   \label{eq:Hxx}
   {\cal H}_{xx}=B_1^x\sigma_1^x + B_2^x\sigma_2^x
	 + J_x\sigma_1^x\sigma_2^x + J_y\sigma_1^y\sigma_2^y + J_z\sigma_1^z\sigma_2^z
	 + J_{yz}\sigma_1^y\sigma_2^z
	 + J_{zy}\sigma_1^z\sigma_2^y.
\end{equation}
The latter can be rewritten in the form
\begin{eqnarray}
   \label{eq:HxxDxGx}
   {\cal H}_{xx}&=&B_1^x\sigma_1^x + B_2^x\sigma_2^x
	 + J_x\sigma_1^x\sigma_2^x + J_y\sigma_1^y\sigma_2^y + J_z\sigma_1^z\sigma_2^z
	 \nonumber\\
	 &+& D_x(\sigma_1^y\sigma_2^z - \sigma_1^z\sigma_2^y)
	 + {\rm \Gamma}_x(\sigma_1^y\sigma_2^z + \sigma_1^z\sigma_2^y),
\end{eqnarray}
where
\begin{equation}
   \label{eq:DxGx}
   D_x=\frac{1}{2}(J_{yz}-J_{zy}),\qquad {\rm\Gamma}_x=\frac{1}{2}(J_{yz}+J_{zy})
\end{equation}
are the $x$-components of Dzyaloshinsky vector and ${\tilde{\rm\Gamma}}$ tensor,
respectively.

So here we have the two-qubit anisotropic Heisenberg spin cluster with $D_x$
and ${\rm\Gamma}_x$ terms of DM and KSEA interactions and additionally in the
nonuniform external fields applied in the transverse $x$-direction.

In Ref.~\cite{Y13} and then in Refs.~\cite{Y14,Y14a}, it has been shown that by means
of double Hadamard transformation $H\otimes H$, where
\begin{equation}
   \label{eq:Had}
   H=\frac{1}{\sqrt{2}}
	 \left(
      \begin{array}{cr}
      1&1\\
      1&-1
      \end{array}
   \right)=H^t
\end{equation}
is the Hadamard transform, any CS matrix $4\times4$ is reduced to the X form (and vice
versa).
Indeed, taking into account relations
\begin{equation}
   \label{eq:HsH}
   H\sigma_x H=\sigma_z,\qquad H\sigma_y H=-\sigma_y,\qquad H\sigma_z H=\sigma_x
\end{equation}
simple calculations yield
\begin{eqnarray}
   \label{eq:HHxxH}
   (H\otimes H){\cal H}_{xx}(H\otimes H)&=&B_1^x\sigma_1^z + B_2^x\sigma_2^z
	 + J_z\sigma_1^x\sigma_2^x + J_y\sigma_1^y\sigma_2^y + J_x\sigma_1^z\sigma_2^z
	 \nonumber\\
	 &+& D_x(\sigma_1^x\sigma_2^y - \sigma_1^y\sigma_2^x)
	 - {\rm \Gamma}_x(\sigma_1^x\sigma_2^y + \sigma_1^y\sigma_2^x).
\end{eqnarray}
Thus, the CS Hamiltonian is returned to the X case up to a reassignment of seven
parameters.

As a result, the discovered remarkable transformation $H\otimes H$ allows to find
correlation functions using the corresponding solutions for the X states.
Knowing the solution for X state we now able to calculate such quantum correlations as
quantum entanglement, discord, and one-way work deficit for the CS case of XYZ model
in the transverse external fields and not only with DM but also with KSEA interaction.

%======================================================================
%\clearpage
%\newpage
\section{
Group-theoretical view on the quantum states
}
\label{sect:GTV}
To find the key to solve the XYZ model with the components $D_y$ and ${\rm\Gamma}_y$
of the DM and KSEA interactions, we analyze the symmetry of the CS matrix using group
theory methods.
In addition, in the future this case will serve us as a heuristic example.

In the most general case, the four-by-four CS matrix is written as
\begin{equation}
   \label{eq:A_CS}
   A_{CS}=
	 \left(
      \begin{array}{cccc}
      A_1&A_2&A_3&A_4\\
      A_5&A_6&A_7&A_8\\
      A_8&A_7&A_6&A_5\\
      A_4&A_3&A_2&A_1
      \end{array}
   \right),
\end{equation}
where the entries are arbitrary real or complex values.
The CS matrix is symmetric about its center.
On the other hand, we may say that CS matrix is such a matrix that commutes with the
operator \cite{Y78}
\begin{equation}
   \label{eq:Uxx}
   U_{xx}=
	 \left(
      \begin{array}{cccc}
      .&.&.&1\\
      .&.&1&.\\
      .&1&.&.\\
      1&.&.&.
      \end{array}
   \right).
\end{equation}
It is clear that $U_{xx}^2$ equals the unity matrix.
One may also claim that the commutativity condition with this operator generates the
CS matrix, i.e., the most general matrix that commutes with $U_{xx}$ is the CS matrix.

Let us find out what consequences the symmetry of matrix (\ref{eq:A_CS}) lead to.
For this purpose we perform a group-theoretical analysis.
The transformation $U_{xx}$ together with the identity transformation $E$
make up the group $\{E,U_{xx}\}$.
This group is second order and has two irreducible representations ${\rm\Gamma}^{(1)}$
and ${\rm\Gamma}^{(2)}$.
The $4\times4$ unit matrix and the matrix (\ref{eq:Uxx}) together give the
original representation $\rm\Gamma$ of this group in the space of the matrix
(\ref{eq:A_CS}).
The characters of $\rm\Gamma$ (traces of representation matrices) equal $\chi(E)=4$
and $\chi(U_{xx})=0$.
Knowing them we can find the multiplicities $a_1$ and $a_2$ with which the
irreducible representations ${\rm\Gamma}^{(1)}$ and ${\rm\Gamma}^{(2)}$, respectively,
are contained in $\rm\Gamma$.
%......................................................................
%                          TABLE 1
\begin{table}[t]
\caption{
Character table of the group $\{E,U_{xx}\}$
\upshape\upshape}
\label{tab:chi}
\begin{tabular}{lcc}
\hline\noalign{\smallskip}
$\{E,U_{xx}\}$ & $E$ & $U_{xx}$ \\
\noalign{\smallskip}\hline\noalign{\smallskip}
%$1.6851637$ & $0.6454108$ & $1.570782\approx90^\circ=89.999^\circ\simeq\pi/2=90^\circ$ \\
${\rm\Gamma}^{(1)}$ & $1$ & $1$ \\
\smallskip
${\rm\Gamma}^{(2)}$ & $1$ & $-1\ \ $ \\
$\rm\Gamma$ & $4$ & $0$ \\
\noalign{\smallskip}\hline
\end{tabular}
\end{table}
%......................................................................
For this purpose it is sufficient to make use of the character table for the group
(Table~\ref{tab:chi}) and the formula \cite{H62}
\begin{equation}
   \label{eq:a_mu}
   a_\mu=\frac{1}{g}\sum_G\chi^{(\mu)}(G)^*\chi(G),
\end{equation}
where $g$ is the order of the group, $\chi^{(\mu)}(G)$ the character of the element
$G$ in the $\mu$-th irreducible representation, and $\chi(G)$ the character of the
same element in the original representation.
Simple calculations yield
\begin{equation}
   \label{eq:a1a2}
   a_1=2,\qquad a_2=2.
\end{equation}
This imply that in the basis where the representation $\rm\Gamma$ of the Abelian
group $\{E,U_{xx}\}$ is completely reducible, the matrix (\ref{eq:A_CS}) will take a
block-diagonal form with two subblocks $2\times2$.

Quasidiagonalizing transformation is constructed from the eigenvectors of the operator
$U_{xx}$ and can be written as
\begin{equation}
   \label{eq:R}
   R=\frac{1}{\sqrt{2}}
	 \left(
      \begin{array}{ccrr}
      1&.&.&1\\
      .&1&1&.\\
      .&1&-1&.\\
      1&.&.&-1
      \end{array}
   \right)=R^t.
\end{equation}
This transformation is orthogonal and symmetric (coincides with its transposition).
After this transformation, the CS matrix (\ref{eq:A_CS}) takes the quasidiagonal form
\begin{equation}
   \label{eq:S-A_CS-S}
   RA_{CS}R=
	 \left(
      \begin{array}{cccc}
      A_1+A_4&A_2+A_3&.&.\\
      A_5+A_8&A_6+A_7&.&.\\
      .&.&A_6-A_7&A_5-A_8\\
      .&.&A_2-A_3&A_1-A_4
      \end{array}
   \right).
\end{equation}
Note that another useful way to practically quasidiagonalize different matrices is to
use for them so-called motion integrals \cite{B64}.

The resulting quasidiagonal form allows it easy to extract all eigenvalues of any CS
matrix and, in particular, of the Hamiltonian and density matrix.
In turn, in some cases, this opens a possibility to direct calculation of quantum
correlations, for example the quantum entanglement of two-qubit quantum CS states
\cite{FKY12}.

Importantly that the matrix $U_{xx}$ can be written as a direct product of Pauli
matrices,
\begin{equation}
   \label{eq:U_xx_dp}
   U_{xx}=\sigma_x\otimes\sigma_x.
\end{equation}
It is this property that allowed to reduce the problem to the known case by applying
the local unitary transformation (double Hadamard transformation) and calculate any
quantum correlations of CS states using the results for the X quantum states.

It is arisen a question either to consider another combinations
$U_{\alpha\beta}=\sigma_\alpha\otimes\sigma_\beta$ ($\alpha,\beta=0, x, y, z$) with
all possible Pauli matrices including the unit matrix $\sigma_0$?
Take, for instance, 
\begin{equation}
   \label{eq:U_zz_dp}
   U_{zz}=\sigma_z\otimes\sigma_z=
	 \left(
      \begin{array}{rrrr}
      1&.&.&.\\
      .&-1&.&.\\
      .&.&-1&.\\
      .&.&.&1
      \end{array}
   \right).
\end{equation}
Simple calculations show that the most general matrix that commutes with $U_{zz}$
has the X form:
\begin{equation}
   \label{eq:A_X}
   A_X=
	 \left(
      \begin{array}{cccc}
      A_1&.&.&A_2\\
      .&A_3&A_4&.\\
      .&A_5&A_6&.\\
      A_7&.&.&A_8
      \end{array}
   \right).
\end{equation}
So we can now give a new definition for the X matrix, namely,
it is such a matrix that commutes with the matrix $U_{zz}=\sigma_z\otimes\sigma_z$
or, in other words, is invariant under the transformations of the group
$\{E,U_{zz}\}$.

In the following sections, we will continue to develop such a group-theoretical
approach.

%======================================================================
%\clearpage
%\newpage
\section{
XYZ chain with $D_y$ and ${\rm\Gamma}_y$ couplings
}
\label{sect:S_Dy}
We return to the consideration of spin systems.
Let us now take the direct product of two $\sigma_y$ matrices,
\begin{equation}
   \label{eq:U_yy_dp}
   U_{yy}=\sigma_y\otimes\sigma_y=
	 \left(
      \begin{array}{rrrr}
      .&.&.&-1\\
      .&.&1&.\\
      .&1&.&.\\
      -1&.&.&.
      \end{array}
   \right),
\end{equation}
and find the most general matrix that commutes with it.
Carrying out the necessary calculations we get the matrix
\begin{equation}
   \label{eq:A_Uyy}
   A_{yy}=
	 \left(
      \begin{array}{rrrr}
      A_1&A_2&A_3&A_4\\
      A_5&A_6&A_7&A_8\\
      -A_8&A_7&A_6&-A_5\\
      A_4&-A_3&-A_2&A_1
      \end{array}
   \right).
\end{equation}
Note that a family of matrices with such a structure is algebraically closed.

Again performing a group-theoretical analysis, as in previous section, we find that
the matrix (\ref{eq:A_Uyy}) can be reduced to a block-diagonal form also with two
sub-blocks of second orders.
The quasidiagonalizing transformation is built from eigenvectors of $U_{yy}$,
Eq.~(\ref{eq:U_yy_dp}), and can be written as
\begin{equation}
   \label{eq:Syy}
   S=\frac{1}{\sqrt{2}}
	 \left(
      \begin{array}{rrrr}
      -1&.&.&1\\
      .&1&1&.\\
      .&1&-1&.\\
      1&.&.&1
      \end{array}
   \right)=S^t.
\end{equation}
Calculations yield
\begin{equation}
   \label{eq:SAyyS}
   SA_{yy}S=
	 \left(
      \begin{array}{cccc}
      A_1-A_4&-A_2-A_3&.&.\\
      -A_5+A_8&A_6+A_7&.&.\\
      .&.&A_6-A_7&A_5+A_8\\
      .&.&A_2-A_3&A_1+A_4
      \end{array}
   \right).
\end{equation}
This results opens a way to extract all eigenvalues of any $A_{yy}$ matrix.

Taking into account hermiticity condition, one can write the density matrix with the
discussed symmetry:
\begin{equation}
   \label{eq:rho_yy}
   \rho_{yy}=
	 \left(
      \begin{array}{lrrc}
      a&\mu&\nu&c\\
      \mu^*&b&d&-\nu^*\\
      \nu^*&d&b&-\mu^*\\
      c&-\nu&-\mu&a
      \end{array}
   \right).
\end{equation}
Similar structure has the Hamiltonian
\begin{equation}
   \label{eq:H_Yyy}
   {\cal H}_{yy}=
	 \left(
      \begin{array}{cccc}
      J_z&-iB_2^y+J_{zx}\ &-iB_1^y+J{xz}&J_x-J_y\\
      iB_2^y+J_{zx}&-J_z&J_x+J_y&-iB_1^y-J_{xz}\\
      iB_1^y+J_{xz}&J_x+J_y&-J_z&-iB_2^y-J_{zx}\\
      J_x-J_y&iB_1^y-J_{xz}\ &iB_2^y-J_{zx}&J_z
      \end{array}
   \right).
\end{equation}
In the Bloch form, this Hamiltonian is given by
\begin{equation}
   \label{eq:Hyy}
   {\cal H}_{yy}=B_1^y\sigma_1^y + B_2^y\sigma_2^y
	 + J_x\sigma_1^x\sigma_2^x + J_y\sigma_1^y\sigma_2^y + J_z\sigma_1^z\sigma_2^z
	 + J_{zx}\sigma_1^z\sigma_2^x
	 + J_{xz}\sigma_1^x\sigma_2^z.
\end{equation}
In terms of the DM and KSEA couplings, this equation is rewritten as
\begin{eqnarray}
   \label{eq:HyyDyGy}
   {\cal H}_{yy}&=&B_1^y\sigma_1^y + B_2^y\sigma_2^y
	 + J_x\sigma_1^x\sigma_2^x + J_y\sigma_1^y\sigma_2^y + J_z\sigma_1^z\sigma_2^z
	 \nonumber\\
	 &+& D_y(\sigma_1^z\sigma_2^x - \sigma_1^x\sigma_2^z)
	 + {\rm \Gamma}_y(\sigma_1^z\sigma_2^x + \sigma_1^x\sigma_2^z),
\end{eqnarray}
where
\begin{equation}
   \label{eq:DyGy}
   D_y=\frac{1}{2}(J_{zx}-J_{xz}),\qquad {\rm\Gamma}_y=\frac{1}{2}(J_{zx}+J_{xz})
\end{equation}
are the $y$-components of Dzyaloshinsky vector and ${\tilde{\rm\Gamma}}$ tensor,
respectively.
So, we come to the completely anisotropic Heisenberg modes in the ``transverse''
external field and with independent $D_y$ and ${\rm\Gamma}_y$ terms of DM and KSEA
interactions.

As already noted above, it is important to find {\em local\/} unitary transformations.
The Hadamard transform diagonalizes the spin matrix $\sigma_x$.
It easy to check that the Pauli matrix $\sigma_y$ is diagonalized by the unitary
transformation
\begin{equation}
   \label{eq:Y}
   Y=\frac{1}{\sqrt2}
	 \left(
      \begin{array}{rr}
      1&1\\
      i&-i
      \end{array}
   \right).
\end{equation}
(This operator can be called a $Y$-transform because it diagonalizes the matrix
$\sigma_y$.)
In the proper representation of the matrix $\sigma_y$\footnote{
One may also choose
$
   \tilde Y=\frac{1}{\sqrt2}
	 \left(
      \begin{array}{rr}
      1&i\\
      i&1
      \end{array}
   \right),
$
that leads to the relations
$\tilde Y^\dagger\sigma_x \tilde Y=\sigma_x$,
$\tilde Y^\dagger\sigma_y \tilde Y=\sigma_z$, and
$\tilde Y^\dagger\sigma_z \tilde Y=-\sigma_y$.
},
\begin{equation}
   \label{eq:YsY}
   Y^\dagger\sigma_x Y=\sigma_y,\quad Y^\dagger\sigma_y Y=\sigma_z,\quad Y^\dagger\sigma_z Y=\sigma_x.
\end{equation}
Double transformation of $Y$ reduces the Hamiltonian ${\cal H}_{yy}$ to
the X form.
Indeed,
\begin{eqnarray}
   \label{eq:YYHyyYY}
   (Y\otimes Y)^\dagger{\cal H}_{yy}(Y\otimes Y)&=&B_1^y\sigma_1^z + B_2^y\sigma_2^z
	 + J_z\sigma_1^x\sigma_2^x + J_x\sigma_1^y\sigma_2^y + J_y\sigma_1^z\sigma_2^z
	 \nonumber\\
	 &+& D_y(\sigma_1^x\sigma_2^y - \sigma_1^y\sigma_2^x)
	 + {\rm \Gamma}_y(\sigma_1^x\sigma_2^y + \sigma_1^y\sigma_2^x).
\end{eqnarray}
The same is valid for the density matrix $\rho_{yy}$: it is also reduced to the X form
by the local unitary transformation consisting of direct product of two $Y$
transforms.

So, we have found a way which allows to calculate the quantum correlations in the XYZ
system with arbitrary components $D_y$ and ${\rm \Gamma}_y$ using the known formulas
for the X states.
At the same time, the way found shows the equivalence of quantum correlation
properties in the system under discussion and in the X (and CS) system.

%======================================================================
%\clearpage
%\newpage
\section{
Classification of quantum states
}
\label{sect:Gen}
The examples considered in the previous sections provide us with a starting point to
explore the invariance under local operations to extend the known results and get new
quantum states.
Consider now the mixed products of spin matrices\footnote{
   We omit the case $U_{00}\ (=\!\!E)$ because $\{E\}$ is the trivial group.
}
\begin{equation}
   \label{eq:Uaa1}
   U_{\alpha\beta}=\sigma_\alpha\otimes\sigma_\beta\qquad (\alpha,\beta=0,x,y,z),
\end{equation}
where $\sigma_0$, $\sigma_x$, $\sigma_y$, and $\sigma_z$ are given by
Eq.~(\ref{eq:s_0xyz}).
The matrices commuting with each given operator $U_{\alpha\beta}$ form algebraically
closed families.
One should be noted that such $U$-operators, up to common coefficients, coincide with
the generators of SU(4) group \cite{R00,ZVSW03,MR19,MR20}.

Repeating calculations similar to Sect.~\ref{sect:GTV}, we find, as above, that the
characters of the initial representation of any group $\{E,U_{\alpha\beta}\}$
are still equal to four and zero, and therefore the multiplicities are again
$a_1=a_2=2$.
As a result, any matrix that commutes with the matrix $U_{\alpha\beta}$ can be reduced
to the block-diagonal form with two subblocks of second order.
Quasidiagonalizing transformations are constructed from the eigenvectors of the
given matrix $U_{\alpha\beta}$.

Finding for each operator $U_{\alpha\beta}$ the matrix originated from the condition
of commutativity and then taking its Hermitian form, we arrive at a collection of
quantum states (and Hamiltonians) which is shown in Table~\ref{tab:15qs}.
%
%......................................................................
%                          TABLE 2
\begin{table}[t]
\caption{
Fifteen quantum states/Hamiltonians generated by $U$-operators and, in braces, spin
%matrices required for their Bloch decompositions
matrix septets required for their Bloch decompositions
\upshape\upshape}
\label{tab:15qs}
\begin{tabular}{cccc}
\noalign{\smallskip}\hline
\hline\noalign{\smallskip}
\smallskip
$$ & $U_{0x}=\sigma_0\otimes\sigma_x$ & $U_{0y}=\sigma_0\otimes\sigma_y$ & $U_{0z}=\sigma_0\otimes\sigma_z$\\
\smallskip
$$&${\cal H}_{0x}, \rho_{0x}\!\!:$&${\cal H}_{0y}, \rho_{0y}\!\!:$&${\cal H}_{0z}, \rho_{0z}\!\!:$\\
\smallskip
$$&$
\left(
   \begin{array}{cccc}
    a&c&\nu&\gamma\\
    c&a&\gamma&\nu\\
    \nu^*&\gamma^*&b&d\\
    \gamma^*&\nu^*&d&b
    \end{array}
\right)
$&$
\left(
   \begin{array}{cccc}
    a&i\mu&\nu&\gamma\\
    -i\mu&a&-\gamma&\nu\\
    \nu^*&-\gamma^*&b&i\delta\\
    \gamma^*&\nu^*&-i\delta&b
    \end{array}
\right)
$&$
\left(
   \begin{array}{cccc}
    a&.&\nu&.\\
    .&b&.&\delta\\
    \nu^*&.&c&.\\
    .&\delta^*&.&d
    \end{array}
\right)
$\\
\smallskip
$$&$\{\sigma_1^x, \sigma_1^y, \sigma_1^z, \sigma_2^x,$&$\{\sigma_1^x, \sigma_1^y, \sigma_1^z, \sigma_2^y,$&$\{\sigma_1^x, \sigma_1^y, \sigma_1^z,  \sigma_2^z,$\\
\smallskip
$$&$\sigma_1^x\sigma_2^x,$&$\sigma_1^y\sigma_2^y,$&$\sigma_1^z\sigma_2^z,$\\
\smallskip
$$&$\sigma_1^z\sigma_2^x,$&$\sigma_1^z\sigma_2^y,$&$\sigma_1^y\sigma_2^z,$\\
$$&$\sigma_1^y\sigma_2^x\}$&$\sigma_1^x\sigma_2^y\}$&$\sigma_1^x\sigma_2^z\}$\\
\noalign{\smallskip}\hline\noalign{\smallskip}
\smallskip
$U_{x0}=\sigma_x\otimes\sigma_0$ & $U_{xx}=\sigma_x\otimes\sigma_x$ & $U_{xy}=\sigma_x\otimes\sigma_y$ & $U_{xz}=\sigma_x\otimes\sigma_z$\\
\smallskip
${\cal H}_{x0}, \rho_{x0}\!\!:$&${\cal H}_{xx}, \rho_{xx}\!\!:$&${\cal H}_{xy}, \rho_{xy}\!\!:$&${\cal H}_{xz}, \rho_{xz}\!\!:$\\
\smallskip
$
\left(
   \begin{array}{cccc}
    a&\mu&c&\gamma\\
    \mu^*&b&\gamma^*&d\\
    c&\gamma&a&\mu\\
    \gamma^*&d&\mu^*&b
    \end{array}
\right)
$&$
\left(
   \begin{array}{cccc}
    a&\mu&\nu&c\\
    \mu^*&b&d&\nu^*\\
    \nu^*&d&b&\mu^*\\
    c&\nu&\mu&a
    \end{array}
\right)
$&$
\left(
   \begin{array}{cccc}
    a&\mu&\nu&i\gamma\\
    \mu^*&b&i\delta&\nu^*\\
    \nu^*&-i\delta&b&-\mu^*\\
    -i\gamma&\nu&-\mu&a
    \end{array}
\right)
$&$
\left(
   \begin{array}{cccc}
    a&\mu&c&\gamma\\
    \mu^*&b&-\gamma^*&d\\
    c&-\gamma&a&-\mu\\
    \gamma^*&d&-\mu^*&b
    \end{array}
\right)
$\\
\smallskip
$\{\sigma_1^x, \sigma_2^x, \sigma_2^y, \sigma_2^z,$&$\{\sigma_1^x, \sigma_2^x,$&$\sigma_1^x, \sigma_2^y,$&$\{\sigma_1^x, \sigma_2^z,$\\
\smallskip
$\sigma_1^x\sigma_2^x,$&$\sigma_1^x\sigma_2^x, \sigma_1^y\sigma_2^y, \sigma_1^z\sigma_2^z,$&$\sigma_1^z\sigma_2^z,$&$\sigma_1^y\sigma_2^y,$\\
\smallskip
$\sigma_1^x\sigma_2^z,$&$\sigma_1^y\sigma_2^z, \sigma_1^z\sigma_2^y\}$&$\sigma_1^x\sigma_2^y, \sigma_1^y\sigma_2^x$&$\sigma_1^z\sigma_2^x,\sigma_1^x\sigma_2^z,$\\
\smallskip
$\sigma_1^x\sigma_2^y\}$&$$&$\sigma_1^y\sigma_2^z,$&$\sigma_1^z\sigma_2^y,$\\
$$&$$&$\sigma_1^z\sigma_2^x\}$&$\sigma_1^y\sigma_2^x\}$\\
\noalign{\smallskip}\hline\noalign{\smallskip}
\smallskip
$U_{y0}=\sigma_y\otimes\sigma_0$ & $U_{yx}=\sigma_y\otimes\sigma_x$ & $U_{yy}=\sigma_y\otimes\sigma_y$ & $U_{yz}=\sigma_y\otimes\sigma_z$\\
\smallskip
${\cal H}_{y0}, \rho_{y0}\!\!:$&${\cal H}_{yx}, \rho_{yx}\!\!:$&${\cal H}_{yy}, \rho_{yy}\!\!:$&${\cal H}_{yz}, \rho_{yz}\!\!:$\\
\smallskip
$
\left(
   \begin{array}{cccc}
    a&\mu&i\nu&\gamma\\
    \mu^*&b&-\gamma^*&i\delta\\
    -i\nu&-\gamma&a&\mu\\
    \gamma^*&-i\delta&\mu^*&b
    \end{array}
\right)
$&$
\left(
   \begin{array}{cccc}
    a&\mu&\nu&i\gamma\\
    \mu^*&b&i\delta&-\nu^*\\
    \nu^*&-i\delta&b&\mu^*\\
    -i\gamma&-\nu&\mu&a
    \end{array}
\right)
$&$
\left(
   \begin{array}{cccc}
    a&\mu&\nu&c\\
    \mu^*&b&d&-\nu^*\\
    \nu^*&d&b&-\mu^*\\
    c&-\nu&-\mu&a
    \end{array}
\right)
$&$
\left(
   \begin{array}{cccc}
    a&\mu&i\nu&\gamma\\
    \mu^*&b&\gamma^*&i\delta\\
    -i\nu&\gamma&a&-\mu\\
    \gamma^*&-i\delta&-\mu^*&b
    \end{array}
\right)
$\\
\smallskip
$\{\sigma_1^y, \sigma_2^x, \sigma_2^y, \sigma_2^z,$&$\{\sigma_1^y, \sigma_2^x,$&$\{\sigma_1^y, \sigma_2^y,$&$\{\sigma_1^y, \sigma_2^z,$\\
\smallskip
$\sigma_1^y\sigma_2^y,$&$\sigma_1^z\sigma_2^z,$&$\sigma_1^x\sigma_2^x, \sigma_1^y\sigma_2^y, \sigma_1^z\sigma_2^z,$&$\sigma_1^x\sigma_2^x,$\\
\smallskip
$\sigma_1^y\sigma_2^z,$&$\sigma_1^x\sigma_2^y, \sigma_1^y\sigma_2^x,$&$\sigma_1^z\sigma_2^x, \sigma_1^x\sigma_2^z\}$&$\sigma_1^y\sigma_2^z,\sigma_1^z\sigma_2^y,$\\
\smallskip
$\sigma_1^y\sigma_2^x\}$&$\sigma_1^z\sigma_2^y,$&$$&$\sigma_1^z\sigma_2^x,$\\
$$&$\sigma_1^x\sigma_2^z\}$&$$&$\sigma_1^x\sigma_2^y\}$\\
\noalign{\smallskip}\hline\noalign{\smallskip}
\smallskip
$U_{z0}=\sigma_z\otimes\sigma_0$ & $U_{zx}=\sigma_z\otimes\sigma_x$ & $U_{zy}=\sigma_z\otimes\sigma_y$ & $U_{zz}=\sigma_z\otimes\sigma_z$\\
\smallskip
${\cal H}_{z0}, \rho_{z0}\!\!:$&${\cal H}_{zx}, \rho_{zx}\!\!:$&${\cal H}_{zy}, \rho_{zy}\!\!:$&${\cal H}_{zz}, \rho_{zz}\!\!:$\\
\smallskip
$
\left(
   \begin{array}{cccc}
    a&\mu&.&.\\
    \mu^*&b&.&.\\
    .&.&c&\nu\\
    .&.&\nu^*&d
    \end{array}
\right)
$&$
\left(
   \begin{array}{cccc}
    a&c&\nu&\gamma\\
    c&a&-\gamma&-\nu\\
    \nu^*&-\gamma^*&b&d\\
    \gamma^*&-\nu^*&d&b
    \end{array}
\right)
$&$
\left(
   \begin{array}{cccc}
    a&i\mu&\nu&\gamma\\
    -i\mu&a&\gamma&-\nu\\
    \nu^*&\gamma^*&b&i\delta\\
    \gamma^*&-\nu^*&-i\delta&b
    \end{array}
\right)
$&$
\left(
   \begin{array}{cccc}
    a&.&.&\gamma\\
    .&b&\delta&.\\
    .&\delta^*&c&.\\
    \gamma^*&.&.&d
    \end{array}
\right)
$\\
\smallskip
$\{\sigma_1^z, \sigma_2^x, \sigma_2^y, \sigma_2^z,$&$\{\sigma_1^z, \sigma_2^x,$&$\{\sigma_1^z, \sigma_2^y,$&$\{\sigma_1^z, \sigma_2^z,$\\
\smallskip
$\sigma_1^z\sigma_2^z,$&$\sigma_1^y\sigma_2^y,$&$\sigma_1^x\sigma_2^x,$&$\sigma_1^x\sigma_2^x, \sigma_1^y\sigma_2^y, \sigma_1^z\sigma_2^z,$\\
\smallskip
$\sigma_1^z\sigma_2^y,$&$\sigma_1^z\sigma_2^x, \sigma_1^x\sigma_2^z,$&$\sigma_1^y\sigma_2^z, \sigma_1^z\sigma_2^y,$&$\sigma_1^x\sigma_2^y,\sigma_1^y\sigma_2^x\}$\\
\smallskip
$\sigma_1^z\sigma_2^x\}$&$\sigma_1^y\sigma_2^z,$&$\sigma_1^x\sigma_2^z,$&$$\\
$$&$\sigma_1^x\sigma_2^y\}$&$\sigma_1^y\sigma_2^x\}$&$$\\
\noalign{\smallskip}\hline
\hline
\end{tabular}
\end{table}
%......................................................................
%
Each quantum state (and hence Hamiltonian) is supplied by a set of Pauli matrices over
which the quantum state or Hamiltonian is decomposed in the form of a linear
combination.

We can look at Table~\ref{tab:15qs} as a four-by-four matrix.
Using this table it is easy to find the Hamiltonians and density matrices of different
spin models.
Let us consider the first row.
The corresponding Hamiltonians are written as
\begin{equation}
   \label{eq:H0x}
   {\cal H}_{0x}={\bf B}_1{\vec \sigma}_1 + B_2^x\sigma_2^x + J_x\sigma_1^x\sigma_2^x
	  +J_{zx}\sigma_1^z\sigma_2^x + J_{yx}\sigma_1^y\sigma_2^x,
\end{equation}
\begin{equation}
   \label{eq:H0y}
   {\cal H}_{0y}={\bf B}_1{\vec \sigma}_1 + B_2^y\sigma_2^y + J_y\sigma_1^y\sigma_2^y
	  +J_{zy}\sigma_1^z\sigma_2^y + J_{xy}\sigma_1^x\sigma_2^y,
\end{equation}
\begin{equation}
   \label{eq:H0z}
   {\cal H}_{0z}={\bf B}_1{\vec \sigma}_1 + B_2^z\sigma_2^z + J_z\sigma_1^z\sigma_2^z
	  +J_{yz}\sigma_1^y\sigma_2^z + J_{xz}\sigma_1^x\sigma_2^z.
\end{equation}
The system (\ref{eq:H0x}) by means of local unitary transform $\sigma_0\otimes H$ and
the system (\ref{eq:H0y}) by the local unitary transformation $\sigma_0\otimes Y$ are
reduced to a structure of the model (\ref{eq:H0z}):
\begin{equation}
   \label{eq:HtoH0z}
   (\sigma_0\otimes H){\cal H}_{0x}(\sigma_0\otimes H)\to{\cal H}_{0z},\quad
   (\sigma_0\otimes Y)\dagger{\cal H}_{0y}(\sigma_0\otimes Y)\to{\cal H}_{0z}.
\end{equation}
Thus, all quantum correlations in these three spin systems are the same.

The mixed members can be rewritten through the DM and KSEA interactions.
For example,
\begin{eqnarray}
   \label{eq:H0zDG}
   {\cal H}_{0z}&=&{\bf B}_1{\vec \sigma}_1 + B_2^z\sigma_2^z + J_z\sigma_1^z\sigma_2^z
	 + D_x(\sigma_1^y\sigma_2^z-\sigma_1^z\sigma_2^y) + {\rm\Gamma}_x(\sigma_1^y\sigma_2^z+\sigma_1^z\sigma_2^y)
	 \nonumber\\
	 &+&D_y(\sigma_1^z\sigma_2^x - \sigma_1^x\sigma_2^z) + {\rm \Gamma}_y(\sigma_1^z\sigma_2^x + \sigma_1^x\sigma_2^z)
\end{eqnarray}
with additional conditions ${\rm \Gamma}_x=D_x$ and  ${\rm \Gamma}_y=-D_y$
in accord with Eqs.~(\ref{eq:DxGx}) and (\ref{eq:DyGy}).
The corresponding density matrix has a characteristic, ``checkerboard'' structure
\begin{equation}
   \label{eq:rho_0z}
    \rho_{0z}=
\left(
   \begin{array}{cccc}
    a&.&\nu&.\\
    .&b&.&\delta\\
    \nu^*&.&c&.\\
    .&\delta^*&.&d
    \end{array}
\right).
\end{equation}
A partial transposition of $\rho_{0z}$, namely $\rho_{0z}^{t_2}$, does not change the
density matrix: $\rho_{0z}^{t_2}=\rho_{0z}$.
Consequently, all eigenvalues stay non-negative and therefore, in accordance with the
positive partial transpose (PPT) criterion
\cite{P96,HHH96}, the state (\ref{eq:rho_0z}) is separable, i.e., its quantum
entanglement (and with it the entanglement of systems with the Hamiltonians
${\cal H}_{0x}$ and ${\cal H}_{0y}$) is identically equal to zero.

Consider now the systems from the first column of Table~\ref{tab:15qs}.
They are
\begin{equation}
   \label{eq:Hx0}
   {\cal H}_{x0}=B_1^x\sigma_1^x + {\bf B}_2{\vec \sigma}_2 + J_x\sigma_1^x\sigma_2^x
	  +J_{xy}\sigma_1^x\sigma_2^y + J_{xz}\sigma_1^x\sigma_2^z,
\end{equation}
\begin{equation}
   \label{eq:Hy0}
   {\cal H}_{y0}=B_1^y\sigma_1^y + {\bf B}_2{\vec \sigma}_2 + J_y\sigma_1^y\sigma_2^y
	  +J_{yz}\sigma_1^y\sigma_2^z + J_{yx}\sigma_1^y\sigma_2^x,
\end{equation}
\begin{equation}
   \label{eq:Hz0}
   {\cal H}_{z0}=B_1^z\sigma_1^z + {\bf B}_2{\vec \sigma}_2 + J_z\sigma_1^z\sigma_2^z
	  +J_{zx}\sigma_1^z\sigma_2^x + J_{zy}\sigma_1^z\sigma_2^y.
\end{equation}
The cross (helical) interactions in these Hamiltonians can also be given in the form
of DM-KSEA interactions.
These models again pass one into another by the local unitary transformations
consisting of the corresponding direct products of operators $H$, $Y$, and $\sigma_0$.
The density matrix corresponding to the Hamiltonian ${\cal H}_{z0}$ has a
block-diagonal (and therefore direct sum) form (see Table~\ref{tab:15qs})
\begin{equation}
   \label{eq:rho_z0}
    \rho_{z0}=
\left(
   \begin{array}{cccc}
    a&\mu&.&.\\
    \mu^*&b&.&.\\
    .&.&c&\nu\\
    .&.&\nu^*&d
    \end{array}
\right)=
\left(
   \begin{array}{cc}
    a&\mu\\
    \mu^*&b
    \end{array}
\right)\oplus
\left(
   \begin{array}{cc}
    c&\nu\\
    \nu^*&d
    \end{array}
\right).
\end{equation}
The quantum entanglement of this state as well as the states corresponding to the
Hamiltonians ${\cal H}_{x0}$ and ${\cal H}_{y0}$ equals zero, again in accordance with
the PPT criterion.

The Hamiltonians (\ref{eq:H0x}) - (\ref{eq:H0z}) are pairwise
connected with the Hamiltonians (\ref{eq:Hx0}) - (\ref{eq:Hz0}) using the spin
exchange operator $P$ introduced by Dirac (see \cite{FLS64}, Sect.~12-2),
\begin{equation}
   \label{eq:P}
   P=\frac{1}{2}(1+\vec\sigma_1\vec\sigma_2)=
	 \left(
      \begin{array}{cccc}
      1&.&.&.\\
      .&.&1&.\\
      .&1&.&.\\
      .&.&.&1
      \end{array}
   \right)=P^t.
\end{equation}
This operator exchanges the first and second qubits ($1\rightleftharpoons2$),
swaps them (similar to the mirror reflection in the plane that separates the qubits):
\begin{equation}
   \label{eq:PsP}
	 P\sigma_\alpha\otimes\sigma_\beta P=\sigma_\beta\otimes\sigma_\alpha,\qquad
	 P\sigma_1^\alpha\sigma_2^\beta P=\sigma_1^\beta\sigma_2^\alpha.
\end{equation}
The matrix (\ref{eq:P}) is an orthogonal transformation that permutes the second and
third rows and columns of any matrix of the fourth order.
One should emphasize that this transformation is not local and therefore, generally
speaking, it changes the value of quantum correlation because the quantum discord and
one-way work deficit depend on which qubit the measurement was performed.
As noted in~\cite{DVB10,MBCPV12}, the discord is not a symmetric quantity,
there are ``left'' and ``right'' discords of the same system.
However if simultaneously with the permutation $P$, the of measured qubit is changed
in the discussed systems then the value of discord (and deficit) will remain
unchanged.
As an example, $P\rho_{0z}P\to\rho_{z0}$ and the ``right'' discord passes to the
``left'' one and vice versa.

Now we turn to the consideration of the ``inner'' part of Table~\ref{tab:15qs}, i.e.,
the systems and their quantum states with the Cartesian indexes $\alpha,\beta=x,y,z$
only.
``Diagonal'' stares ($\rho_{xx}$, $\rho_{yy}$, and $\rho_{zz}$) have already been
discussed in detail in previous sections.
All of them are related by local unitary transformations and therefore the quantum
correlations are the same and can be calculated using formulas available for the X
state.

The ``off-diagonal'' Hamiltonians from the upper triangle part of the table are
written as follows
\begin{equation}
   \label{eq:Hxy-a}
   {\cal H}_{xy}
	 =B_1^x\sigma_1^x + B_2^y\sigma_2^y	+ J_z\sigma_1^z\sigma_2^z
	 +J_{xy}\sigma_1^x\sigma_2^y + J_{yx}\sigma_1^y\sigma_2^x
   +J_{yz}\sigma_1^y\sigma_2^z
	 +J_{zx}\sigma_1^z\sigma_2^x,
\end{equation}
\begin{equation}
   \label{eq:Hxz}
   {\cal H}_{xz}
	 =B_1^x\sigma_1^x + B_2^z\sigma_2^z	+ J_y\sigma_1^y\sigma_2^y
	 +J_{zx}\sigma_1^z\sigma_2^x + J_{xz}\sigma_1^x\sigma_2^z
   +J_{zy}\sigma_1^z\sigma_2^y
	 +J_{yx}\sigma_1^y\sigma_2^x,
\end{equation}
\begin{equation}
   \label{eq:Hyz}
   {\cal H}_{yz}
	 =B_1^y\sigma_1^y + B_2^z\sigma_2^z	+ J_x\sigma_1^x\sigma_2^x
	 +J_{zy}\sigma_1^z\sigma_2^y + J_{yz}\sigma_1^y\sigma_2^z
   +J_{zx}\sigma_1^z\sigma_2^x
	 +J_{xy}\sigma_1^x\sigma_2^y.
\end{equation}
These Hamiltonians can be expressed via DM-KSEA interactions.
For instance,
\begin{equation}
   \label{eq:Hxy-}
   {\cal H}_{xy}
	 =B_1^x\sigma_1^x + B_2^y\sigma_2^y	+ J_z\sigma_1^z\sigma_2^z
   +{\vec D}\!\cdot\!({\vec\sigma}_1\times{\vec\sigma}_2)
   +{\vec\sigma}_1\!\cdot\!{\tilde{\rm\Gamma}}\!\cdot\!{\vec\sigma}_2
\end{equation}
with conditions ${\rm\Gamma}_x=D_x$ and ${\rm\Gamma}_y=D_y$ whereas ${\rm\Gamma}_z$
and $D_z$ are arbitrary independent quantities.
The Hamiltonians of a lower part of the ``off-diagonal'' systems are
\begin{equation}
   \label{eq:Hyx}
   {\cal H}_{yx}
	 =B_1^y\sigma_1^y + B_2^x\sigma_2^x	+ J_z\sigma_1^z\sigma_2^z
	 +J_{xy}\sigma_1^x\sigma_2^y + J_{yx}\sigma_1^y\sigma_2^x
   +J_{zy}\sigma_1^z\sigma_2^y
	 +J_{xz}\sigma_1^x\sigma_2^z,
\end{equation}
\begin{equation}
   \label{eq:Hzx}
   {\cal H}_{zx}
	 =B_1^z\sigma_1^z + B_2^x\sigma_2^x	+ J_y\sigma_1^y\sigma_2^y
	 +J_{zx}\sigma_1^z\sigma_2^x + J_{xz}\sigma_1^x\sigma_2^z
   +J_{yz}\sigma_1^y\sigma_2^z
	 +J_{xy}\sigma_1^x\sigma_2^y,
\end{equation}
\begin{equation}
   \label{eq:Hzy}
   {\cal H}_{zy}
	 =B_1^z\sigma_1^z + B_2^y\sigma_2^y	+ J_x\sigma_1^x\sigma_2^x
	 +J_{zy}\sigma_1^z\sigma_2^y + J_{yz}\sigma_1^y\sigma_2^z
   +J_{yx}\sigma_1^y\sigma_2^x
	 +J_{xz}\sigma_1^x\sigma_2^z.
\end{equation}
Remarkably that all these six ``off-diagonal'' Hamiltonians (\ref{eq:Hxy-a}) -
(\ref{eq:Hyz}) and (\ref{eq:Hyx}) - (\ref{eq:Hzy}) are reduced to the Hamiltonian of
the X model by local unitary transformations composed of the operators $H$, $Y$, and
$\sigma_0$.
Indeed, 
\begin{equation}
   \label{eq:HtoH}
   (H\otimes Y)^\dagger{\cal H}_{xy}(H\otimes Y)\to{\cal H}_{zz},\quad
   (H\otimes\sigma_0){\cal H}_{xz}(H\otimes\sigma_0)\to{\cal H}_{zz}.
\end{equation}
Similarly for other cases.

So, out of fifteen quantum states, nine ($\rho_{\alpha\beta}$ with
$\alpha,\beta=x,y,z$) are transformed among themselves by local unitary
transformations.
Their quantum correlations are complete identical to each other and are calculated by
the formulas for the X state.
 
The quantum states of the six remaining models are separable and therefore without
quantum entanglement.
These models consist of two equal subclasses $\rho_{0\alpha}$ and $\rho_{\alpha0}$
($\alpha=x,y,z$) in each of which the quantum discord and other quantum correlations
are equivalent to each other since the states are connected via local unitary
transformations.
Moreover, the states from different subclasses are paired trough the spin exchange
transformation $P$ thanks to which the ``right'' quantum discord of one member of a
pair equals the ``left'' discord of other member of the same pair.
Unfortunately, among both subclasses there is no one quantum state for which the
quantum discord is known.
However, most recently Zhou, Hu, and Jing \cite{ZHJ20} have evaluated the ``right''
and ``left'' quantum discords for the quantum sate
$$\rho=\frac{1}{4}(1+\vec s_1\vec\sigma_1+c_3\sigma_1^z\sigma_2^z)$$
(see Theorem 2.3 in their paper \cite{ZHJ20}).
It would be interesting to extend this result to the general ``checkerboard''
($\rho_{0z}$) or 
two-block-$(2\times2)$-diagonal 
%``block-diagonal'' 
($\rho_{z0}$) quantum states.

%======================================================================
\clearpage
\newpage
\section{Results and perspectives}
\label{sect:Concl}
We have analyzed fifteen types of two-spin systems in an external field, with the
exchange bounds, and with indirect interactions occurring through the orbital magnetic
moments.
The structures of Hamiltonians and density matrices are presented in an obvious form
(Table~\ref{tab:15qs}).

\begin{figure}[b]
\begin{center}
\epsfig{file=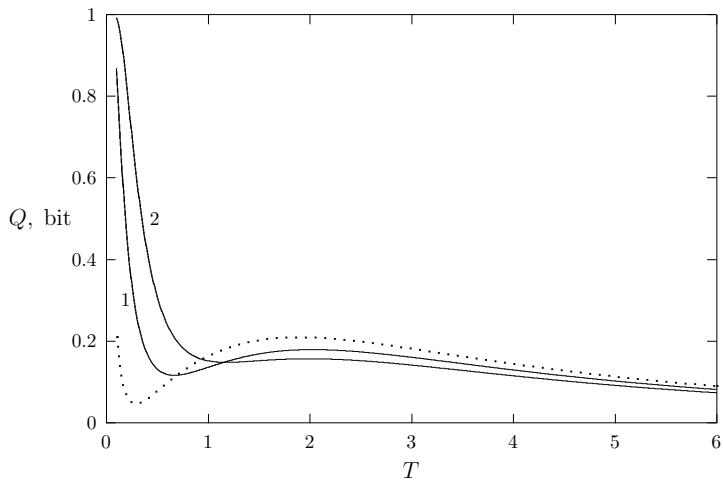,width=9.4cm}
\caption{
Quantum discord $Q$ vs temperature $T$ for the model (\ref{eq:HXzz}) by $B_1^z=B_2^z=0$,
$J_x=-1$, $J_y=-1.5$, $J_z=-2$, $D_z=1.8$,
and ${\rm\Gamma}_z=0$~(dotted line), $0.3$~(solid line 1), $0.5$~(solid line 2)
}
\label{fig:zq1805}
\end{center}
\end{figure}

The originality and new feature of our results in comparison with other works,
\cite{Z07,LWC08,KJL09,LC09,CY10,TAMAP13,Z14,P19,SMG20}, is that we take into account
not only the DM interactions, but also the KSEA ones.
The latter interactions make their own changes in the behavior of quantum
correlations.
For example, in recent work \cite{P19} a local minimum was found due to DM couplings
(black line in Fig.~6c of Ref.~\cite{P19}).
This behavior is reproduced by the dotted line in Fig.~\ref{fig:zq1805}.
We also performed calculations for non-zero values of the constant ${\rm\Gamma}_z$
shown by curves 1 and 2 in Fig.~\ref{fig:zq1805}.
It is seen that the KSEA interactions suppress the local minimum of the quantum
discord.
%......................................................................
%                          FIGURE 1
%......................................................................

We have classified fifteen types of quantum states, each of which contains seven
parameters.
A detailed study of the behavior of quantum correlations in them will require a
separate extensive work in the future.

We have also shown that, from viewpoint of quantum correlation properties, all systems
are divided into two groups: the systems with the X quantum states (up to local
unitary transformations) for which the developed theory for the calculation of quantum
correlations is available and systems with checkerboard-like or block-diagonal non-X
quantum states in which the quantum entanglement is absent whereas the question about
the quantum discord and other quantum correlations remains open.

%======================================================================
%\section*{Acknowledgment}
\vspace{-10mm}
\section*{}
{\bf Acknowledgment}\ This work was performed as a part of the state task of the RF,
%State Registration No.~0089-2019-0002,
CITIS \# AAAA-A19-119071190017-7.

%======================================================================
%\clearpage
%\newpage

%======================================================================

\end{document}